\documentclass[12pt, a4paper]{article}
\usepackage{a4wide,amssymb,amsmath,amscd}
\usepackage{color}

\begin{document}

\topmargin -2pt


\headheight 0pt

\topskip 0mm \addtolength{\baselineskip}{0.20\baselineskip}

\vspace{5mm}

\begin{center}
{\Large \bf Friedmann equation and the emergence of cosmic space} \\

\vspace{10mm}

{\sc Ee Chang-Young}\footnote{cylee@sejong.ac.kr} and
{\sc Daeho Lee}\footnote{dhleep@gmail.com}
\\

\vspace{1mm}

{\it Department of Physics and Institute of Fundamental Physics,\\
    Sejong University, Seoul 143-747, Korea}\\

\vspace{10mm}
{\bf ABSTRACT}
\end{center}


\noindent
\noindent

In this paper, we show that Padmanabhan's conjecture for the emergence of cosmic space [arXiv:1206.4916]
 holds for the flat Friedmann-Robertson-Walker universe in Einstein gravity
 but does not hold for the non-flat case
 unless one uses the aerial volume instead of the proper volume.
 Doing this, we also show that various works extending Padmanabhan's conjecture to non-Einstein and
 non-flat cases have serious shortfalls.
This analysis is done using the Friedmann equation with the further assumptions of the holographic principle
and the equipartition rule of energy.

\vfill

\thispagestyle{empty}

\newpage




\section{Introduction}

The discovery of the thermodynamic properties of black holes in the 1970s implied a close connection between gravity and thermodynamics~\cite{bek73,haw7576,bch73}.
It also motivated the idea that gravity could be an emergent phenomenon.
Based on these observations, Jacobson~\cite{jac95} showed that the Einstein equation can be derived from
 black hole thermodynamics applied for a local accelerating observer.
Verlinde then suggested that gravity is an entropic force induced by the change of entropy associated with position~\cite{verl11}.
He derived Newton's force by applying the holographic principle and equipartition rule of energy.

Padmanabhan~\cite{pad12} recently proposed a governing equation that would
 control the expansion of cosmic space.
It has been known that the universe of pure de Sitter type satisfies the holographic equipartition,
$N_{\texttt{sur}}=N_{\texttt{bulk}}$, where $N_{\texttt{sur}}$ and $N_{\texttt{bulk}}$ are the
degrees of freedom~(DOF) of the boundary surface and of the bulk, respectively.
Under the assumption that our universe is asymptotically de Sitter,
Padmanabhan considered that the expansion of the universe is being driven towards holographic equipartition.
Thus, he conjectured that the governing equation for the expansion of cosmic space
 is given by
\begin{equation}
\label{Padm_conj}
\frac{dV}{dt}=L_p^2(N_{\texttt{sur}}-N_{\texttt{bulk}}),
\end{equation}
where $V$ is the volume of cosmic space enclosed by the apparent horizon $\tilde{r}_A$, and $L_p$ is the Planck length.
Using the above relation and with further assumptions of
 the holographic principle and the equipartition rule of energy, he
  succeeded in obtaining the Friedmann equation of
the (3+1)-dimensional flat Friedmann-Robertson-Walker (FRW) universe in Einstein gravity.

Then in Ref.~\cite{Cai12}, Cai obtained the Friedmann equations
for an (n+1)-dimensional flat FRW universe in the Einstein, Gauss-Bonnet, and Lovelock gravity cases using Padmanabhan's conjecture. However, in the Gauss-Bonnet and Lovelock cases, Cai used  an effective volume for the volume change but used
 the plain ordinary volume for the bulk DOF.

In order to avoid this discrepancy,
 the plain ordinary volume was used both for the volume change and for the bulk DOF  in Ref.~\cite{ylw12}.
But it was not free, and in order to obtain the Friedmann equation,
Padmanabhan's  relation~(\ref{Padm_conj}) had to be severely modified in both Gauss-Bonnet and Lovelock cases.

The application to the non-flat case was first done in Ref.~\cite{she13}.
Although the Friedmann equation was obtained with a slight modification of Padmanabhan's conjecture,
this work was criticized for using the aerial volume instead of
using the proper invariant volume for non-flat geometry.

Ref.~\cite{ek13} tried to make up for this shortcoming by using the proper invariant volume
in the non-flat Einstein case.
However, they also had to severely modify Padmanabhan's relation in order to obtain the Friedmann equation.
In that work `the effective Planck length' was newly introduced solely to contain all the complications of the
time dependence replacing the proportionality factor, so that
the original form of Padmanabhan's conjecture could be maintained.
But this is certainly not commensurate with Padmanabhan's conjecture, wherein
the rate of volume change should be just proportional to $ N_{\texttt{sur}}-N_{\texttt{bulk}} $.

In this paper, we show that all these conjectured relations appearing in Refs.~\cite{pad12,Cai12,ylw12,she13,ek13}
 simply follow from the Friedmann equation when we
 assume the holographic principle and the equipartition rule of energy.
The organization of the paper is as follows.
In section 2, we show that Padmanabhan's relation and its modified versions can be obtained from the Friedmann equation
for the flat FRW universe in the Einstein, Gauss-Bonnet, and Lovelock gravity cases.
In section 3, we perform the same analysis for the non-flat case in (3+1)-dimensions.
We conclude with a discussion in section 4.
Throughout the paper, we use the natural units $k_B=c=\hbar=1$  for the sake of brevity.
\\

\section{Emergence of cosmic space for a flat FRW universe}

In this section, we show that the Friedmann equation of the flat FRW universe
yields Padmanabhan's relation when the holographic principle and the equipartition rule of energy are assumed.

 For later reference,  we consider an (n+1)-dimensional FRW universe with the metric
\begin{equation}
\label{frw}
ds^2=-dt^2+a^2(t)\left(\frac{dr^2}{1-kr^2 }+r^2 d\Omega_{n-1}^2 \right),
\end{equation}
where $d\Omega_{n-1}^2$ denotes the line element of the $(n-1)$-dimensional unit sphere.
Here, the spatial curvature constant $k$ corresponds to a closed, flat and open universe for $k=+1, 0,$
and $ -1$, respectively. The metric (\ref{frw}) can be rewritten as
\begin{equation}
\label{frw_re}
ds^2=h_{ab} dx^a dx^b +\tilde{r}^2 d\Omega_{n-1}^2,~a,b=0,1,
\end{equation}
where $\tilde{r}=a(t)r$, $h_{ab}=\texttt{diag}(-1,a^2/1-kr^2)$, and $(x^0,x^1)=(t,r)$.
The apparent horizon in (\ref{frw_re}) is defined as the marginally trapped surface with vanishing expansion and is determined by the relation $h^{ab}\partial_a \tilde{r}\partial_b \tilde{r}=0$. Thus, the radius of the apparent horizon is given by
\begin{equation}
\label{apphorizon}
\tilde{r}_A= \frac{1}{\sqrt{H^2+k/a^2}},
\end{equation}
where $H \equiv \dot{a}/a$ is the Hubble parameter.
The Hawking temperature associated with the apparent horizon is given by
\begin{equation}
\label{hawtem}
T_H= \frac{1}{2\pi \tilde{r}_A}.
\end{equation}
In this paper, we will consider the case in which the distribution of matter and energy takes the form of a perfect fluid.
Then the Friedmann equations in the (n+1)-dimensional Einstein gravity are given by~\cite{Caikim05}:
\begin{equation}
\label{friedeq1_einf}
H^2+\frac{k}{a^2}=\frac{16\pi L_{p}^{n-1}}{n(n-1)}\rho,
\end{equation}
\begin{equation}
\label{friedeq2_einf}
\frac{\ddot{a}}{a}=-\frac{8\pi L_{p}^{n-1}}{n(n-1)}[(n-2)\rho+np].
\end{equation}
We now restrict ourselves to the flat case ($k=0$).
 Since the volume enclosed by the apparent horizon $\tilde{r}_A$ in the flat case is given by
$V=\Omega_n \tilde{r}_{A}^n$, where $\Omega_n$ is the volume of the unit $n$-sphere,
 the rate of volume change is given by
\begin{eqnarray}
\label{vcflat}
\frac{dV}{dt}= n\Omega_n \tilde{r}_A^{n-1}\dot{\tilde{r}}_A.
\end{eqnarray}
Then with the use of the Friedmann equation~(\ref{friedeq2_einf}),  it is given by
\begin{eqnarray}
\label{vcapp_einf}
\frac{dV}{dt}= L_p^{n-1}  \left(\frac{A}{ L_p^{n-1}}+\frac{8\pi\tilde{r}_A V }{n-1}[(n-2)\rho+np]\right),
\end{eqnarray}
where
$A=n\Omega_n \tilde{r}_A^{n-1}$.
The bulk Komar energy in an (n+1)-dimensional flat spacetime is given by~\cite{cco10}:
\begin{eqnarray}
\label{komar}
E=\frac{[(n-2)\rho+np]}{(n-2)}V.
\end{eqnarray}
With the use of the equipartition rule of energy, the bulk DOF is given by
\begin{eqnarray}
\label{ndofobf}
N_{\texttt{bulk}}= \frac{2|E|}{T_H}=
-4\pi \tilde{r}_A V \frac{[(n-2)\rho+np]}{n-2}.
\end{eqnarray}
Note that $(n-2)\rho+np<0$, since $N_{\texttt{bulk}}>0$. \\
The surface DOF can be identified as in Ref.~\cite{verl11} by
\begin{eqnarray}
\label{ndofos_einstein}
N_{\texttt{sur}}=\alpha \frac{A}{L_p^{n-1}},
\end{eqnarray}
where $\alpha=(n-1)/2(n-2)$.
The inclusion of the coefficient $\alpha$ is necessary to attain the correct identification
with the Newton constant in the higher dimensional case~\cite{verl11}.
Eq.~(\ref{vcapp_einf}) can then be written as
\begin{eqnarray}
\label{vcapp_ein}
\frac{dV}{dt}=
\tilde{L}_p^{n-1}(N_{\texttt{sur}}-N_{\texttt{bulk}}),
\end{eqnarray}
where $\tilde{L}_p$ is defined by
 \[
 \tilde{L}_p^{n-1} \equiv L_p^{n-1}/\alpha.
 \]
This is the relation that Cai used in Ref.~\cite{Cai12} to derive the Friedmann equation
in the (n+1)-dimensional flat Einstein case.
 Note that
in the (3+1)-dimensional case, $\alpha=1$ and  $\tilde{L}_p = L_p$, and so
we recover the relation (\ref{Padm_conj}) that Padmanabhan conjectured.

In the remaining part of this section, we will perform the same analysis in the Gauss-Bonnet and Lovelock gravity cases.

 First, we consider the Gauss-Bonnet case.
Recall that the apparent horizon and volume are given by the same formulas as in the Einstein case.
%
%
The Friedmann equations for the Gauss-Bonnet gravity are given by~\cite{Caikim05}:
\begin{equation}
\label{friedeq1_GBf}
H^2+\frac{k}{a^2}+\tilde{\alpha}\left(H^2+\frac{k}{a^2}\right)^2=\frac{16\pi L_{p}^{n-1}}{n(n-1)}\rho,
\end{equation}
\begin{eqnarray}
\label{friedeq2_GBf}
\left(\dot{H}-\frac{k}{a^2}\right)\left[1+2\tilde{\alpha}\left(H^2+\frac{k}{a^2}\right)\right]
&+& \left(H^2+\frac{k}{a^2}\right)\left[1+ \tilde{\alpha}\left(H^2+\frac{k}{a^2}\right)\right]
\nonumber \\
&=&-\frac{8\pi L_{p}^{n-1}}{n(n-1)}[(n-2)\rho+np],
\end{eqnarray}
where $\tilde{\alpha}=(n-2)(n-3)\alpha$.

Now we get into the flat case ($k=0$).
The rate of the volume change \eqref{vcflat} is given by
\begin{eqnarray}
\label{vcapp_GBf}
\frac{dV}{dt} = \frac{L_{p}^{n-1}}{(1+2\tilde{\alpha}\tilde{r}_A^{-2})}
             \left( \frac{A(1+\tilde{\alpha}\tilde{r}_A^{-2})}{L_{p}^{n-1}}+\frac{8\pi\tilde{r}_A V }{n-1}[(n-2)\rho+np]\right).
\end{eqnarray}
In the above we used the Friedmann equation (\ref{friedeq2_GBf}) with $k=0$.
The bulk DOF is given by the same formula  as in the Einstein case:
\begin{eqnarray}
\label{ndofobf_GB}
N_{\texttt{bulk}}^{GB}=-4\pi \tilde{r}_A V \frac{[(n-2)\rho+np]}{n-2}.
\end{eqnarray}
If we use the same ansatz for the surface DOF as in  Ref.~\cite{Cai12},
\begin{eqnarray}
\label{ndofos_GB}
N_{\texttt{sur}}^{GB}=\frac{A(1+\tilde{\alpha}\tilde{r}_A^{-2})}{\tilde{L}_p^{n-1}},
\end{eqnarray}
then   Eq.~(\ref{vcapp_GBf}) can be expressed as
\begin{eqnarray}
\label{fvcapp_GBfne}
\frac{dV}{dt}
=\frac{\tilde{L}_p^{n-1}}{(1+2\tilde{\alpha}\tilde{r}_A^{-2})} (N_{\texttt{sur}}^{GB}-N_{\texttt{bulk}}^{GB}).
\end{eqnarray}
This is still different from Padmanabhan's relation~(\ref{Padm_conj}) by the  factor $(1+2\tilde{\alpha}\tilde{r}_A^{-2})^{-1}$.
To deal with this, Cai \cite{Cai12} introduced the effective volume
related to the effective area,
 which has been used to deal with the entropy of black holes in the Gauss-Bonnet gravity case:
%
\begin{eqnarray}
\label{effarea}
\tilde{A}=A\left(1+\frac{n-1}{n-3} 2\tilde{\alpha}\tilde{r}_A^{-2}\right).
\end{eqnarray}
The effective volume can be obtained using the relation
$d\tilde{V}/d\tilde{A}=\tilde{r}_A/(n-1)$.
In this way, we get the relation
%
\begin{eqnarray}
\label{effvcapp_GBf}
\frac{d\tilde{V}}{dt}=\tilde{L}_p^{n-1}(N_{\texttt{sur}}^{GB}-N_{\texttt{bulk}}^{GB}),
\end{eqnarray}
which was used in Ref.~\cite{Cai12} to get the Friedmann equation in  the Gauss-Bonnet case.
However, there is a catch in this derivation;
in the above equation, on the right-hand side the plain ordinary volume was used for the bulk DOF \eqref{ndofobf_GB},
 but on the left-hand side the effective volume was used to calculate the rate of volume change.

In order to avoid the above discrepancy, the authors of Ref.~\cite{ylw12} assumed a relation
modified severely
from the original Padmanabhan conjecture (\ref{Padm_conj}).
However, their modified relation can be also obtained from the Friedmann equation, as we see now.
They used the same bulk and surface DOFs as in the flat Einstein case, Eqs.~(\ref{ndofobf}) and (\ref{ndofos_einstein}).
Then one can show that the rate of volume change
(\ref{vcapp_GBf}) can be written as
\begin{eqnarray}
\label{fvcapp_GBylwf}
\frac{dV}{dt}
=L_p^{n-1} \frac{ (N_{\texttt{sur}}-N_{\texttt{bulk}})/\alpha + \tilde{\alpha}K(N_{\texttt{sur}}/\alpha)^{1+\frac{2}{1-n}}
}{1+2\tilde{\alpha} K(N_{\texttt{sur}}/\alpha)^{\frac{2}{1-n}}
},
\end{eqnarray}
where $K=(n\Omega_n / L_p^{n-1})^{2/(n-1)}$.
In fact, Eq.~(\ref{fvcapp_GBylwf}) is the modified relation that was assumed in Ref.~\cite{ylw12} to derive
the Friedmann equation in the Gauss-Bonnet case.
\\

Next we consider the Lovelock gravity case.
As in the  Gauss-Bonnet case, the spacetime is described by the same metric (\ref{frw}).
The apparent horizon and the volume remain the same as in the  Gauss-Bonnet  case.
The Friedmann equations in the Lovelock gravity are given by~\cite{Caikim05}:
\begin{equation}
\label{friedeq1_Lovef}
\sum_{i=1}^{m}\hat{c}_i\left(H^2+\frac{k}{a^2}\right)^i=\frac{16\pi L_{p}^{n-1}}{n(n-1)}\rho,
\end{equation}
\begin{eqnarray}
\label{friedeq2_Lovef}
\left(\dot{H}-\frac{k}{a^2}\right)\sum_{i=1}^{m}i\hat{c}_i\left(H^2+\frac{k}{a^2}\right)^{i-1}
+\sum_{i=1}^{m}\hat{c}_i\left(H^2+\frac{k}{a^2}\right)^i
=-\frac{8\pi L_{p}^{n-1}}{n(n-1)}[(n-2)\rho+np],
\end{eqnarray}
where $m=[n/2]$ and the coefficients are given by
\begin{equation}
\label{c_i}
\hat{c}_1=1,~\hat{c}_i=c_i \prod_{j=3}^{2m}(n+1-j)~~\texttt{for}~~ i>1.
\end{equation}
Once again we consider the flat case ($k=0$).
Then with the use of the Friedmann equation~(\ref{friedeq2_Lovef}) with $k=0$,
 the rate of volume change \eqref{vcflat}  is given by
\begin{eqnarray}
\label{vcapp_Lovef}
\frac{dV}{dt}= \frac{L_{p}^{n-1}}{\sum_{i=1}^m i\hat{c}_i \tilde{r}_A^{2(1-i)}}
             \left(\frac{A\sum_{i=1}^m \hat{c}_i\tilde{r}_A^{2(1-i)}}{L_{p}^{n-1}}+\frac{8\pi\tilde{r}_A V }{(n-1)}[(n-2)\rho+np] \right).
\end{eqnarray}
As in the Gauss-Bonnet case, the bulk DOF is the same as in the Einstein  case,
\begin{eqnarray}
\label{ndofobf_Love}
N_{\texttt{bulk}}^{L}=-4\pi \tilde{r}_A V \frac{[(n-2)\rho+np]}{n-2},
\end{eqnarray}
and if we use the same ansatz for the surface DOF as in  Ref.~\cite{Cai12},
\begin{eqnarray}
\label{ndofos_Love}
N_{\texttt{sur}}^L= \frac{A}{\tilde{L}_p^{n-1}}\sum_{i=1}^m \hat{c}_i \tilde{r}_A^{2(1-i)},
\end{eqnarray}
then Eq.~(\ref{vcapp_Lovef}) can be written as
\begin{eqnarray}
\label{fvcapp_LovefV}
\frac{dV}{dt} = \frac{\tilde{L}_p^{n-1}}{\sum_{i=1}^m i\hat{c}_i \tilde{r}_A^{2(1-i)}} (N_{\texttt{sur}}^L-N_{\texttt{bulk}}^L).
\end{eqnarray}

Now, we introduce the effective volume
related to the effective area,
 which has been used to deal with the entropy of black holes in the Lovelock case, as  Cai did in \cite{Cai12},
\begin{eqnarray}
\label{effareal}
\tilde{A}=A\sum_{i=1}^m \frac{i(n-1)}{(n-2i+1)}\hat{c}_i \tilde{r}_A^{2(1-i)},
\end{eqnarray}
and use the relation $d\tilde{V}/d\tilde{A}=\tilde{r}_A/(n-1)$,
then  Eq.~(\ref{fvcapp_LovefV}) can be expressed as
\begin{eqnarray}
\label{effvcapp_Lovef}
\frac{d\tilde{V}}{dt}= \tilde{L}_p^{n-1}(N_{\texttt{sur}}^L-N_{\texttt{bulk}}^L).
\end{eqnarray}
This is the relation used in Ref.~\cite{Cai12} to get the Friedmann equation in  the Lovelock case.
However, this derivation has the same problem as in the Gauss-Bonnet case;
on the right-hand side the plain ordinary volume was used for the bulk DOF (\ref{ndofobf_Love}),
 but on the left-hand side the effective volume was used to calculate the rate of volume change.

To avoid this discrepancy, the authors of Ref.~\cite{ylw12} did the same thing as in the Gauss-Bonnet case;
they used the same bulk and surface DOFs as in the flat Einstein case, Eqs.~(\ref{ndofobf}) and (\ref{ndofos_einstein}).
Then one can easily show that the rate of volume change
(\ref{vcapp_Lovef}) can be written as
\begin{eqnarray}
\label{fvcapp_Lylwf}
\frac{dV}{dt}
=L_p^{n-1} \frac{(N_{\texttt{sur}}-N_{\texttt{bulk}})/\alpha+
\sum_{i=2}^m \tilde{c}_iK_i(N_{\texttt{sur}}/\alpha)^{1+ \frac{2(i-1)}{1-n}
}
 }{1+\sum_{i=2}^m i \tilde{c}_i K_i (N_{\texttt{sur}}/\alpha)^{ \frac{2(i-1)}{1-n}
 }},
\end{eqnarray}
where $K_i=(n\Omega_n /L_p^{n-1})^{2(i-1)/(n-1)} $.
This is indeed the modified relation used in Ref.~\cite{ylw12} to derive the Friedmann equation in the Lovelock case.
\\

\section{Emergence of cosmic space for a non-flat FRW universe}

In this section, we check Padmanabhan's conjecture in the non-flat case.
For the sake of brevity, we do this in the (3+1)-dimensional  case.

The invariant volume of the space enclosed by the apparent horizon $\tilde{r}_A$ for the (3+1)-dimensional
non-flat FRW universe is given by
\begin{eqnarray}
\label{inv_nonflat}
V_k &=& 4\pi a^3 \int_{0}^{\tilde{r}_A/a} \frac{r^{2}}{\sqrt{1-kr^2}}dr,
\end{eqnarray}
where $k=\pm 1$.
Note that $V_k$ becomes $4\pi \tilde{r}_A^3/3$ in the limit $k \rightarrow 0$.

Sheykhi was the first to apply  Padmanabhan's conjecture to the non-flat case~\cite{she13}.
However, instead of using the proper invariant volume \eqref{inv_nonflat},
Sheykhi used the aerial volume for the flat case, $V=\Omega_n \tilde{r}_{A}^n$.
Thus, the rate of volume change
was given by Eq.~\eqref{vcflat}.
In the Einstein case, using the Friedmann equation~(\ref{friedeq2_einf}) and
the relation \eqref{apphorizon}, the rate of volume change is given by the right-hand side (RHS)
of Eq.~\eqref{vcapp_einf}  multiplied by a factor  $H\tilde{r_A}$.
Adopting the same definitions of the bulk and surface DOFs as in the flat case,
Eqs.~(\ref{ndofobf}) and (\ref{ndofos_einstein}),
one can easily check that the rate of volume change is given by
\begin{equation}
\label{Padm_conjsh}
\frac{dV}{dt}=\tilde{L}_p^{n-1} H\tilde{r_A} (N_{\texttt{sur}}-N_{\texttt{bulk}}),
\end{equation}
where $V=\Omega_n \tilde{r}_A^{n}$.
This is just the modified relation that Sheykhi assumed
in order to obtain the Friedmann equation in the non-flat Einstein case.

In the Gauss-Bonnet and Lovelock  cases, Sheykhi followed Cai's method~\cite{Cai12}.

Using the same volume  as in the Einstein case and with the aid of the Friedmann equations (\ref{friedeq2_GBf})
and (\ref{friedeq2_Lovef}),
 the rates of volume change in the Gauss-Bonnet  and Lovelock cases are given by the RHS of
 Eqs. \eqref{vcapp_GBf} and  \eqref{vcapp_Lovef},  respectively, with the same additional multiplication factor, $H\tilde{r_A}$.

Then with the same definitions of the bulk and surface DOFs as Cai's in the flat case,
one can check that the rate of the change of the effective volume can be written as
\begin{equation}
\label{Padm_conjGBsh}
\frac{d\tilde{V}}{dt}=\tilde{L}_p^{n-1} H\tilde{r_A} (N_{\texttt{sur}}^{GB}-N_{\texttt{bulk}}^{GB})
\end{equation}
in the Gauss-Bonnet case, and
\begin{equation}
\label{Padm_conjLLsh}
\frac{d\tilde{V}}{dt}=\tilde{L}_p^{n-1} H\tilde{r_A} (N_{\texttt{sur}}^{L}-N_{\texttt{bulk}}^{L})
\end{equation}
in the Lovelock case.
These two relations are what Sheykhi assumed for the derivations of the Friedmann equations
 in the Gauss-Bonnet and Lovelock  cases.
Obviously these derivations have the same problem as Cai's, namely using the
effective volume for the volume change but using the ordinary volume for the  bulk  DOF.

Now we look into how the Friedmann equation can yield the modified Padmanabhan's relation
 assumed in Ref.~\cite{ek13}
when the proper volume is used.

By applying the Friedmann equation~(\ref{friedeq2_einf}) and
with the use of the relation~\eqref{inv_nonflat}, the rate of the change of the invariant volume is given by
\begin{eqnarray}
\label{inv_t}
\frac{dV_k}{dt} &=&  4\pi\tilde{r}_A^2 \left(\frac{\dot{\tilde{r}}_A}{H\tilde{r}_A}-1+H\tilde{r}_A \frac{V_k}{\bar{V}_k}\right)  \nonumber \\
 &=& L_p^2 \Big[ \frac{A}{L_p^2}H\tilde{r}_A\frac{V_k}{\bar{V}_k}
 +\frac{\bar{V}_k}{V_k} 4\pi\tilde{r}_A(\rho+3p)V_k\Big]
\end{eqnarray}
where $\bar{V}_k= 4\pi\tilde{r}_A^3/3$ and $A=4\pi\tilde{r}_A^2$.

Since the bulk Komar energy in the non-flat case is given by~\cite{cco10}
\begin{eqnarray}
\label{komar}
E_k=(\rho+3p)V_k,
\end{eqnarray}
 the bulk DOF  with the  assumption of the equipartition rule of energy is given by
\begin{eqnarray}
\label{ndofob3_einstein}
N_{\texttt{bulk}}=\frac{2|E_k|}{T_H}=
-4\pi\tilde{r}_A(\rho+3p)V_k.
\end{eqnarray}
Here $\rho+3p<0$, since $N_{\texttt{bulk}}>0$.
 Applying the holographic principle, the surface DOF is given by
\begin{eqnarray}
\label{ndofos3_einstein}
N_{\texttt{sur}}=A/L_p^2.
\end{eqnarray}
Now, the rate of volume change (\ref{inv_t}) can be written as
\begin{eqnarray}
\label{vcapp_einn}
\frac{dV_k}{dt}=L_p^{2}\left(H \tilde{r}_A\frac{V_k}{\bar{V}_k}N_{\texttt{sur}}
-\frac{\bar{V}_k}{V_k}N_{\texttt{bulk}}\right) \equiv L_p^{2}\Delta \mathcal{N}.
\end{eqnarray}
Here, we introduce $\Delta \mathcal{N}$ for  later use.

Obviously the above result shows that Padmanabhan's conjecture does not hold
 in the non-flat case if one uses the proper invariant volume.
%
%
Thus in Ref.~\cite{ek13}, in order to conform with the original form of Padmanabhan's conjecture,
the Planck length had to be replaced with
 `the  effective Planck length,' which is a complicated function of time.
Now, we briefly look into how `the  effective Planck length' of Ref.~\cite{ek13} can be obtained.

 From Eqs.~(\ref{ndofob3_einstein}) and (\ref{ndofos3_einstein}) and using the Friedmann equation~(\ref{friedeq2_einf}),
one can write the following relation:
\begin{eqnarray}
\label{diffN1}
N_{\texttt{sur}}-N_{\texttt{bulk}} =
\frac{4\pi\tilde{r}_A^{2}}{L_p^2}\frac{V_k}{\bar{V}_k}
\left[ \left(\frac{\dot{\tilde{r}}_A}{H\tilde{r}_A}-1+\frac{\bar{V}_k}{V_k}\right)
\right] \equiv \Delta N.
\end{eqnarray}
Using the above relation, one can now rewrite Eq.~\eqref{vcapp_einn} in the following form:
\begin{eqnarray}
\label{vcapp_einnp}
\frac{dV_k}{dt} \equiv  L_p^2 f_k(t)\Delta N,
\end{eqnarray}
where
\begin{eqnarray}
f_k(t) \equiv \frac{\Delta \mathcal{N}} {\Delta N} = \frac{L_p^2}{4\pi\tilde{r}_A^{2}}\frac{\bar{V}_k}{V_k}
\frac{\left(H \tilde{r}_A\frac{V_k}{\bar{V}_k}N_{\texttt{sur}}
       -\frac{\bar{V}_k}{V_k}N_{\texttt{bulk}}\right)}
{\left(\frac{\dot{\tilde{r}}_A}{H\tilde{r}_A}-1+\frac{\bar{V}_k}{V_k}\right)}.
\label{fkt}
\end{eqnarray}
Relation \eqref{vcapp_einnp} is  what was assumed in Ref.~\cite{ek13}
to derive the Friedmann equation.
 $\sqrt{L_p^2 f_k(t)}$  was dubbed there as `the effective Planck length'.
However, the complicated time dependence of the function $f_k(t)$ just tells us that
the rate of volume change is not proportional to $\Delta N= N_{\texttt{sur}}-N_{\texttt{bulk}} $,
in stark contrast to Padmanabhan's idea that it should be.
%
\\

\section{Conclusion}

In this paper we have shown that how Padmanabhan's original conjecture on the evolution of cosmic space
and its modified versions
in various cases can be obtained from the Friedmann equation.
In doing this, we have assumed the holographic principle and the equipartition rule of energy
 as they were assumed by Padmanabhan and others who used `the conjecture' to obtain the Friedmann equation.

Padmanabhan's original relation emerges without difficulty
 from the Friedmann equation  in the flat Einstein case.
However, in the non-flat Einstein case, the Friedmann equation
emerges only
when one uses the aerial volume
instead of the proper  volume.
Furthermore, in the Gauss-Bonnet and Lovelock cases,
Padmanabhan's conjecture
has to be severely modified
even for the flat FRW universe to obtain the Friedmann equation.

In the non-flat Einstein case, Sheykhi~\cite{she13} used a relatively simple modification
 of Padmanabhan's conjecture to obtain the Friedmann equation.
However,  Sheykhi used the aerial volume  instead of the proper invariant volume
 for non-flat space.
Ref.~\cite{ek13} tried to make up for this shortcoming with the use of the proper invariant volume.
However, Padmanabhan's conjecture had to be  severely modified to
such an extent that it looses the meaning of the original conjecture.

 For the Gauss-Bonnet and Lovelock gravity theories,  Cai~\cite{Cai12} first applied
 Padmanabhan's conjecture to obtain the Friedmann equation in the flat case.
He considered the effective volume instead of the plain volume to calculate
the rate of volume change and  managed to obtain the Friedmann equation
with the original Padmanabhan relation.
However, there was a catch in his derivations; he used the plain volume for
the bulk DOF instead of the effective volume.
In order to remedy this inconsistency, the plain volume was used both for the volume change and for the bulk DOF in Ref.~\cite{ylw12}.
However, Padmanabhan's conjecture had to be severely modified in \cite{ylw12}
in order to get the Friedmann equations in both Gauss-Bonnet and Lovelock  cases.

The analysis carried out in this paper shows that Padmanabhan's conjecture can be obtained from
 the Friedmann equation
without any difficulty
in the flat Einstein case.
For the non-flat case, Padmanabhan's conjecture with a slight modification can be obtained from
the Friedmann equation only when one uses the aerial volume.
However, applying it to non-Einstein gravity cases, one encounters serious difficulties and
the conjecture itself
has to be  heavily modified  to  get the Friedmann equation.

Padmanabhan's assumption that the expansion of the universe is being driven
towards holographic equipartition
is seemingly very attractive.
Nonetheless, the work in this paper shows that
it is only compatible with the aerial volume not with the proper volume in
 the non-flat Einstein case. Furthermore, it also has severe difficulties
in non-Einstein cases.

However, if we assume that Padmanabhan's conjecture is correct, then
the above limitation or trouble other than the flat Einstein case
might give us some hint as to
 why our universe is spatially flat.
\\

%

\section*{Acknowledgments}

This work was supported by the National Research Foundation (NRF) of Korea Grant funded by
the Korean Government, NRF-2011-0025517.
\\


\end{document}